\documentstyle[seceq,epsf]{ptptex}

\begin{document}

\newcommand{\beq}{\begin{equation}}
\newcommand{\eeq}{\end{equation}}
\newcommand{\beqn}{\begin{eqnarray}}
\newcommand{\eeqn}{\end{eqnarray}}
\newcommand{\pa}{\partial}
\newcommand{\vp}{\varphi}
\def\bI{\hbox{$\,I\!\!\!\!-$}}

\begin{center}
{\large\bf{Fully general relativistic simulation of 
merging binary clusters\\
-- Spatial gauge condition --
}}
~\\
~\\
Masaru Shibata \\
~\\
{\em Department of Physics, 
University of Illinois at Urbana-Champaign, Urbana, IL 61801, USA \\
{\rm and}\\
Department of Earth and Space Science,~Graduate School of
  Science,~Osaka University,~Toyonaka, Osaka 560-0043, Japan 
}\\
\end{center}
~\\
\begin{abstract}
~\\
We have carried out simulations of the coalescence 
between two relativistic 
clusters of collisionless particles using a 3D numerical relativity 
code. We have adopted a new spatial gauge condition obtained by 
slightly modifying the 
minimum distortion gauge condition proposed by Smarr and York 
and resulting in a simpler equation for the shift vector. 
Using this gauge condition, 
we have performed several simulations of the merger between two 
identical clusters in which we have varied 
the compaction, the type of internal motion in the 
clusters, and the magnitude of the orbital velocity. 
As a result of the coalescence, either a new rotating cluster 
or a black hole is formed. 
In the case in which a black hole is not formed, 
simulations could be carried out for a time much longer
than the dynamical time scale, and the resulting 
gravitational waveforms were calculated fairly accurately: 
In these cases, 
the amplitude of gravitational waves emitted can be 
$\sim 10^{-18}(M/10^6M_{\odot})$ at a distance 4000Mpc, and 
$\sim 0.5\%$ of the rest mass energy may be dissipated by 
the gravitational wave emission in the final phase of the merger. 
These results confirm that the new spatial gauge condition 
is promising in many problems at least 
up to the formation of black holes. 
In the case in which a black hole is formed, on the other hand, 
the gauge condition seems to be less adequate, but we 
suggest a strategy to improve it in this case. 
All of the results obtained confirm the robustness of 
our formulation and the ability of our code for 
stable evolution of strong gravitational fields of 
compact binaries. 
\end{abstract}
%\pacs{04.25.Nx}
%\vskip2pc]

\section{Introduction}

The coalescences of relativistic binaries are the most promising sources 
for kilometer-size laser interferometers such as 
LIGO,\cite{LIGO,KIP} VIRGO,\cite{VIRGO} GEO,\cite{GEO} and 
TAMA,\cite{TAMA} which will be in operation in the next five years, 
as well as for future space-based interferometric 
gravitational wave detectors such as LISA.\cite{LISA} 
When a signal of gravitational waves from such 
compact objects is detected, it will be analyzed using 
matched filter techniques, and 
a variety of astrophysical information will be extracted.\cite{KIP} 
In order to apply this technique, 
theoretical templates of gravitational waveforms will be needed, and 
this motivates the recent theoretical study of gravitational waves 
emitted from coalescing binaries. 
The post-Newtonian approximation has been shown to be a 
powerful tool 
for the study of the inspiraling phase, 
and a large effort devoted to this study has produced many 
satisfactory results.\cite{blanchet} However, 
in order to study the final phase of the coalescence, 
in which strong general relativistic effects dominate, 
no approximation of general relativity is valid, 
and 3D numerical relativistic simulations appear to be 
the only promising approach. 

Although a few results have been obtained 
recently,\cite{review} 
there are still a number of fundamental questions that need 
to be addressed. 
Among them, it is particularly important and urgent 
to find appropriate gauge conditions which allow one to 
perform stable simulations and to extract gravitational 
waves accurately. 
As shown by many numerical simulations, 
the maximal slice condition seems to be an adequate time slice 
condition at least up to the formation of black holes. 
However, a similarly reliable spatial gauge condition in 3D 
numerical relativity has not yet been studied sufficiently. 
More than 20 years ago, Smarr and York \cite{smarr} proposed 
a minimum distortion (MD) gauge condition, 
which is expected to have several excellent mathematical properties. 
From a numerical point of view, however, 
the MD gauge condition requires the solution of 
a complicated vector elliptic equation, and this can be a  
very time-consuming task. 

Their MD gauge condition was contrived from the idea that the 
global change rate of the conformal spatial metric should be 
minimized.\cite{smarr} 
This is a rather restrictive condition, and we believe that 
a gauge condition in which the global 
change rate is not {\it exactly} minimized but instead 
only {\it approximately} minimized can prove to be appropriate, 
as long as the geometrical distortion of the three space 
does not increase quickly. Thus, in this paper, 
we propose an approximate 
minimum distortion (AMD) gauge condition which 
is similar to Smarr-York's MD gauge condition and in which 
the global change rate is expected to be only approximately 
minimized. In a numerical simulation, the AMD gauge condition 
is much more easily imposed than the original 
MD gauge condition. 

In a previous paper,\cite{gw3d} we 
presented results from numerical simulations of 
black hole formation in which 
collisionless particles were used as matter sources of the 
Einstein equation. That study 
demonstrated that our formulation is robust, 
that it allows for 
stable numerical evolution, and that it can be 
applied to problems such as the formation of 
black holes from collision of two clusters, or 
gravitational collapse. In this paper, 
we use our code to simulate the merger between two identical 
binary clusters, 
adopting an AMD gauge condition, and we demonstrate that 
this formulation with a new spatial gauge condition 
allows for stable numerical evolutions of merging 
binaries and for fairly accurate calculation of gravitational waves. 

Numerical simulations of the merger between relativistic clusters may 
become a source of important information in 
astrophysics and gravitational wave astronomy. 
It has been in fact proposed 
that supermassive black holes in galactic centers might be 
formed through complex phenomena such as 
gravitational collapse and bar instability, involving 
the coalescence between clusters of collisionless 
compact stars.\cite{rees} If such phenomena 
occurred frequently in the early 
universe, they could have produced low frequency 
gravitational waves which would be detected by the planned 
interferometric detectors in space.\cite{LISA} 
In particular, the coalescence of two compact clusters 
could be an important source. 
The simulation presented in this paper may play a role 
in predicting the amplitude and frequency of 
the gravitational waves emitted during such coalescences. 

The paper is organized as follows. In \S 2, 
we present the basic equations to be solved in the 
numerical simulation. 
In \S 3, we describe the spatial 
AMD gauge condition adopted in this paper and 
briefly discuss its properties. 
In \S 4, we describe the methods employed in order 
to analyze the gravitational 
waves radiated. In \S 5, after we briefly 
describe the method to set the initial conditions, 
we present numerical results 
for simulations of the coalescence of two clusters. 
Section 6 is devoted to summary. 
Throughout this paper, we adopt units in which $G=1=c$. 
Latin and Greek indices denote spatial components ($1-3$) 
and spacetime components ($0-3$), respectively. 
As spatial coordinates, we use the Cartesian coordinates 
$x^k=(x, y, z)$ with $r=\sqrt{x^2+y^2+z^2}$. 

\section{Basic equations}

Our formulation of the Einstein equation is 
described in detail in previous papers.\cite{SN,gw3d} 
We here briefly recall the basic equations and refer the reader 
to the above cited works for details. 

We write the line element in the form 
\beq
ds^2=(-\alpha^2+\beta_k\beta^k)dt^2
+2\beta_i dx^i dt+\gamma_{ij}dx^i dx^j ,
\eeq
where $\alpha$, $\beta^i~(\beta_i=\gamma_{ij}\beta^j)$ 
and $\gamma_{ij}$ are the lapse function, 
shift vector, and 3D spatial metric, respectively. 
We define the quantities
\beqn
&& \gamma={\rm det}(\gamma_{ij}) \equiv e^{12\phi},\\
&& \tilde \gamma_{ij} \equiv e^{-4\phi}\gamma_{ij}, 
~{\rm i.e.},~{\rm det}(\tilde \gamma_{ij})=1, \\
&& \tilde A_{ij} \equiv 
e^{-4\phi} \Bigl(K_{ij}-{1 \over 3} \gamma_{ij} K \Bigr),
\eeqn
where $K_{ij}$ is the extrinsic curvature, and $K=K_k^{~k}$. 
The indices of $\tilde A_{ij}$ and/or $\tilde A^{ij}$ are 
raised and lowered in terms of $\tilde \gamma_{ij}$ and 
$\tilde \gamma^{ij}$. Numerical computations are carried out 
using $\tilde \gamma_{ij}$, $\tilde A_{ij}$, $\phi$ and $K$  
as dynamical variables rather than $\gamma_{ij}$ and $K_{ij}$. 
Hereafter, we use $D_i$ and $\tilde D_i$ as the covariant 
derivatives with respect to $\gamma_{ij}$ and 
$\tilde \gamma_{ij}$, respectively. 

Since our matter source is represented by 
collisionless particles, 
its energy momentum tensor is written as   
\beq
T_{\mu\nu}=m_p \sum_{a=1}^N 
{\delta^{(3)}(x^j-x^j_a) \over \alpha e^{6\phi}} 
\biggl( {u_{\mu} u_{\nu} \over u^0 }\biggr)_a,
\eeq
where $m_p$ is the rest mass of each particle 
(we assume that all particles have the same rest mass), 
$x^j_a$ denotes the position of the $a$-th particle, 
$N$ is the total particle number, 
$u_{\mu}$ is the four-velocity of a particle, and 
$\delta^{(3)}(x^j-x^j_a)$ is the Dirac delta function in 
the 3D spatial 
hypersurface. Note that the rest mass density $\rho_*$ and 
the total rest mass (i.e., conserved mass) $M_*$ are written as 
\beqn
&& \rho_*=m_p \sum_{a=1}^N \delta^{(3)}(x^j-x^j_a),\\
&& M_*=4Nm_p, \label{mast}
\eeqn
where the factor $4$ in Eq. (\ref{mast}) 
arises because in this paper we assume a plane symmetry 
with respect to the equatorial plane $(z=0)$ as well as a 
$\pi$-rotation symmetry around the $z$-axis, and 
treat only a quadrant region (see below). 

The equations of motion for the particles are derived 
from the geodesic equations and are written in the form 
\beq
{d u_i \over dt} = -\alpha u^0 \alpha_{,i}+u_j \beta^j_{~,i}
-{u_j u_k \over 2 u^0} \gamma^{jk}_{~~,i}~~~~,\label{eq1}
\eeq
where $Q_{,j}=\pa_j Q=\pa Q /\pa x^j$. 
Once $u_i$ is obtained, 
$u^0$ is determined from the normalization relation of the 
four-velocity as 
\beq
(\alpha u^0)^2=1+\gamma^{ij} u_i u_j~~. 
\eeq
We also have a relation between a coordinate velocity 
$dx^i/dt\equiv u^i/u^0$ and $u_j$ as  
\beq
{dx^i \over dt}= -\beta^i + {\gamma^{ij} u_j \over u^0}~~. 
\label{eq2}
\eeq
Equations (\ref{eq1}) and (\ref{eq2}) are the basic 
equations for evolution of the particle's position and velocity. 

As customary in a 3+1 decomposition, 
the Einstein equation is split into the constraint 
and evolution equations. 
The Hamiltonian and momentum constraint equations are 
given by 
\beqn
&& R- \tilde A_{ij} \tilde A^{ij}+{2 \over 3} K^2=16\pi E,
\label{ham}\\
&& D_i \tilde A^i_{~j}-{2 \over 3}D_j K=8\pi J_j, \label{mom}
\eeqn
where 
\beqn
&& E=m_p \sum_{a=1}^N (u^0)_a \alpha e^{-6\phi} 
\delta^{(3)}(x^k-x^k_a),\\
&& J_i=m_p \sum_{a=1}^N (u_i)_a e^{-6\phi} \delta^{(3)}(x^k-x^k_a),
\eeqn
and $R$ is the scalar curvature with respect to $\gamma_{ij}$. 
We solve these constraint equations only when setting 
initial conditions. 

The evolution equations for the geometric variables 
$\tilde \gamma_{ij}$, $\phi$, $\tilde A_{ij}$ and $K$ are given by 
\cite{gw3d}
\beqn
&&(\pa_t - \beta^k \pa_k) \tilde \gamma_{ij}
=-2\alpha \tilde A_{ij} 
+\tilde \gamma_{ik} \beta^k_{~,j}+\tilde \gamma_{jk} \beta^k_{~,i}
-{2 \over 3}\tilde \gamma_{ij} \beta^k_{~,k}~~, \label{heq} \\
%%%%%%%%%%%%
&&(\pa_t - \beta^k \pa_k) \tilde A_{ij} 
= e^{ -4\phi } \biggl[ \alpha \Bigl(R_{ij}
-{1 \over 3}\gamma_{ij} R \Bigr) 
-\Bigl( D_i D_j \alpha - {1 \over 3}\gamma_{ij} D_k D^k \alpha \Bigr)
\biggr] \nonumber \\
&& \hskip 2.8cm 
+\alpha (K \tilde A_{ij} - 2 \tilde A_{ik} \tilde A_j^{~k})
+\beta^k_{~,i} \tilde A_{kj}+\beta^k_{~,j} \tilde A_{ki}
-{2 \over 3} \beta^k_{~,k} \tilde A_{ij} \nonumber \\
&&\hskip 2.8cm 
-8\pi\alpha e^{-4\phi} \Bigl( S_{ij}-{1 \over 3} \gamma_{ij} S_k^{~k}
\Bigr), \label{aijeq} \\
%%%%%%%%%%%%
&&(\pa_t - \beta^k \pa_k) \phi = {1 \over 6}\Bigl( 
-\alpha K + \beta^k_{~,k} \Bigr), \label{peq} \\
%%%%%%%%%%%
&&(\pa_t - \beta^k \pa_k) K
=\alpha \Bigl(\tilde A_{ij} \tilde A^{ij}+{1 \over 3}K^2 \Bigr)
-D_k D^k \alpha +4\pi \alpha (E+ S_k^{~k}), \label{keq}
\eeqn
where $R_{ij}$ is the Ricci tensor with respect to $\gamma_{ij}$, 
and 
\beq
S_{ij}=m_p \sum_{a=1}^N  \biggl({u_i u_j \over u^0 }\biggr)_a
{\delta^{(3)}(x^k-x^k_a) \over \alpha e^{6\phi} }. 
\eeq
We also solve the 
following equation for $F_i \equiv \delta^{jk} 
\tilde \gamma_{ij,k}$: \cite{SN,gw3d,Nakamura}
\beqn
(\pa_t - \beta^k \pa_k)F_i&& = 2\alpha \Bigl(f^{kj} \tilde A_{ik,j}
+f^{kj}_{~~,j} \tilde A_{ik}-{1 \over 2} \tilde A^{jl} h_{jl,i} 
+6\phi_{,k} \tilde A^k_{~i}-{2\over 3}K_{,i} \Bigr) \nonumber \\
&&~
-2\delta^{jk} \alpha_{,k} \tilde A_{ij} 
+ \delta^{jl} \beta^k_{~,l}h_{ij,k}
+(\tilde \gamma_{il}\beta^l_{~,j}+\tilde \gamma_{jl}\beta^l_{~,i}
-{2\over 3}\tilde \gamma_{ij} \beta^l_{~,l})_{,k}\delta^{jk} 
\nonumber \\
&&~-16\pi \alpha J_i~~. \label{fijeq}
\eeqn
Here $\delta_{ij}$ denotes the Kronecker delta, 
$h_{ij}=\tilde \gamma_{ij}-\delta_{ij}$, and 
$f^{ij}=\tilde \gamma^{ij}-\delta^{ij}$. 
$F_i$ is used to compute $R_{ij}$ and $R$.\cite{gw3d} 
As argued in previous papers,\cite{SN,gw3d} 
the introduction of $F_i$ is necessary for 
stable numerical computation. 

As the time slice condition, we 
use an approximate maximal slice condition. 
Namely, we 
determine $\alpha$ by demanding 
that the right-hand side of 
Eq. (\ref{keq}) becomes approximately zero, 
using a strategy discussed in a previous paper.\cite{gw3d} 
A discussion of the spatial gauge condition 
used in this paper will be presented in the next section. 

The numerical methods employed in solving Eqs. 
(\ref{eq1}), (\ref{eq2}), (\ref{heq})--(\ref{keq}) and 
(\ref{fijeq}), and 
for finding the apparent horizon are almost the same as 
those discussed in a previous paper,\cite{gw3d} except 
for the following two differences. 
The first difference is due to the appearance 
of the transport terms such as 
$-\beta^k\pa_k \tilde \gamma_{ij}$ in the evolution 
equations of the geometric variables. This is 
because we choose a non-zero $\beta^k$ 
(in a previous paper \cite{gw3d} we adopted $\beta^k=0$). 
A discussion of our strategy for handling 
such transport terms is described in Appendix A. 
The second difference is due to the different underlying 
symmetries assumed here. 
In this paper, we solve equations 
in the quadrant region $-L \leq x \leq L$, $0 \leq y, z \leq L$, where 
$L$ denotes the location of the outer boundaries, 
assuming $\pi$-rotation 
symmetry around the $z$-axis as well as 
the plane symmetry with respect to the $z=0$ plane 
(in a previous paper\cite{gw3d} 
we assumed triplane symmetry). 
We note that we perform simulations of binary clusters whose 
mass centers orbit in the $z=0$ plane (see \S 5). 
Boundary conditions in the $y=0$ plane are as follows:
\beqn
Q(x,0,z)&=&Q(-x,0,z),\\
Q^A(x,0,z)&=&-Q^A(-x,0,z),~~~Q_A(x,0,z)=-Q_A(-x,0,z),\\
Q^z(x,0,z)&=&Q^z(-x,0,z),~~~\hskip 4mm Q_z(x,0,z)=Q_z(-x,0,z),\\
Q_{AB}(x,0,z)&=&Q_{AB}(-x,0,z),\\
Q_{Az}(x,0,z)&=&-Q_{Az}(-x,0,z),\\
Q_{zz}(x,0,z)&=&Q_{zz}(-x,0,z),
\eeqn
where $A=x$ or $y$, and $Q$, $Q^i({\rm or}~Q_i)$ and $Q_{ij}$ 
denote arbitrary scalar, vector and tensor quantities, respectively. 
Note that the boundary conditions at the outer boundaries
are the same as those in a previous paper \cite{gw3d} 
except for that of $F_i$ for which we impose $F_i=O(r^{-3})$ 
in this paper. It is also convenient 
to define the quantities
\beqn
&&j(r)=m_p \sum_{a=1}^N (x u_y-y u_x)_a ~~{\rm for}~ r_a < r,\\
&&m_*(r)=m_p N_p(r) ,
\eeqn
where $r_a=\sqrt{x_a^2 + y_a^2 + z_a^2}$ and $N_p(r)$ denotes 
the number of particles which are in a radius $r$. 
We consider $j(r)$ to be 
the approximate $z$-component of the angular momentum within $r$,
while $m_*(r)$ denotes the total rest mass within $r$. 

\section{Spatial gauge condition}

The MD gauge condition proposed by Smarr and York \cite{smarr} 
can be written as 
\beq
D^i (e^{4\phi} \pa_t {\tilde \gamma_{ij}} )=0,
\eeq
or, equivalently, as  
\beq
\tilde D^i (e^{6\phi} \pa_t {\tilde \gamma_{ij}} )=0. \label{SMMD}
\eeq
More explicitly, their MD gauge condition reduces to an equation 
for $\beta^k$: 
\beqn
&&\tilde D^i \Bigl(\tilde D_i \tilde \beta_j+\tilde D_j 
\tilde \beta_i
-{2 \over 3}\tilde \gamma_{ij} \tilde D_k \tilde \beta^k \Bigr)
+6 \tilde D^i \phi 
\Bigl(\tilde D_i \tilde \beta_j+\tilde D_j \tilde \beta_i
-{2 \over 3}\tilde \gamma_{ij} \tilde D_k \tilde \beta^k \Bigr) 
\nonumber \\
&&\hskip 2cm -2 \tilde A_{ ij} \tilde D^i \alpha 
-{4 \over 3}\alpha \tilde D_j K 
=16\pi\alpha J_j.\label{mdeq1}
\eeqn
Here $\tilde \beta_k=\tilde \gamma_{kl}\beta^l$ and 
$\tilde \beta^k=\beta^k$. 
Smarr-York's MD gauge condition is also derived by 
minimizing the following action $I$ with respect to  $\beta^k$ 
on three spacelike hypersurfaces:\cite{smarr} 
\beq
I=\int d^3x (\pa_t {\tilde \gamma_{ij}}) 
(\pa_t {\tilde \gamma_{kl}})
\tilde \gamma^{ik} \tilde \gamma^{jl} e^{6\phi}. 
\eeq 
Thus, by choosing the Smarr-York's MD 
gauge condition, the global change rate of 
$\tilde \gamma_{ij}$ {\it based on the action $I$} 
is minimized in every three hypersurface.

Since $I$ does not have a special physical meaning,  
we may consider an alternative gauge condition by 
slightly changing the definition of the action. 
In particular, we define another action $I'$ as 
\beq
I'=\int d^3x (\pa_t {\tilde \gamma_{ij}}) 
(\pa_t {\tilde \gamma_{kl}})
\tilde \gamma^{ik} \tilde \gamma^{jl} .
\eeq
This corresponds to defining the action in the conformal three 
space in which the determinant of the metric 
det$(\tilde \gamma_{ij})$ is unity. In this case, 
by taking the variation of $I'$ with respect to $\beta^k$, 
we obtain a different MD gauge condition, 
\beq
\tilde D^i (\pa_t {\tilde \gamma_{ij}})=0,\label{MDnew}
\eeq
or more explicitly, 
\beqn
&&\tilde D^i \Bigl(\tilde D_i \tilde \beta_j
+\tilde D_j \tilde \beta_i
-{2 \over 3}\tilde \gamma_{ij} \tilde D_k \tilde \beta^k \Bigr)
-2 \tilde A_{ ij} (\tilde D^i \alpha - 6\alpha \tilde D^i \phi)
\nonumber \\
&& \hskip 2cm -{4 \over 3}\alpha \tilde D_j K 
=16\pi\alpha J_j.\label{md2}
\eeqn
A merit of this condition may be that 
we do not have a coupling term between $\tilde \beta^k$ 
and $\phi$, and hence the equation for 
$\tilde \beta^k$ is slightly simplified compared with 
Eq. (\ref{mdeq1}). However, it is still complicated to solve 
in numerical computation. 

Although it is desirable to adopt a shift vector in which 
$I$ or $I'$ is {\it exactly} minimized, we believe that 
we do not always have to use such 
a shift vector. We may probably use another shift vector in which 
$I$ or $I'$ is {\it approximately} minimized. 
Using this idea, we derive a gauge condition by slightly 
changing Eq. (\ref{md2}): We rewrite 
the covariant derivative operator $\tilde D_i$ 
acting on $\tilde \beta^k$ as a partial derivative; i.e., 
we solve the following equation to determine 
$\beta^k (=\tilde \beta^k)$: 
\beq
\delta_{ij} \Delta \beta^i + {1 \over 3} \beta^k_{~,kj}
-2\tilde A_{ ij} (\tilde D^i \alpha - 6\alpha \tilde D^i \phi) 
-{4 \over 3}\alpha \tilde D_j K 
 =16\pi\alpha J_j.\label{amdeq}
\eeq
Here $\Delta $ is the Laplacian in the flat space. 
In this case, Eq. (\ref{amdeq}) can be rewritten into 
simple elliptic equations for a vector $P_i$ and a 
scalar $\eta$ using the transformation 
\beq
\beta^j=\delta^{ji}\biggl[
{7 \over 8}P_i - {1 \over 8}(\eta_{,i}+P_{k,i} x^k) \biggr],
\eeq
where $P_i$ and $\eta$ satisfy 
\beqn
&&\Delta P_i = S_i, \\
&&\Delta \eta= -S_i x^i, 
\eeqn
and 
\beq
S_i\equiv 16\pi\alpha J_i 
+2\tilde A_{ ij} (\tilde D^j \alpha - 6\alpha \tilde D^j \phi)
+{4 \over 3}\alpha \tilde D_i K. 
\eeq
In this way, the equation for $\beta^k$ is 
significantly simplified and reduces to equations similar to 
those we have solved in the initial value problem \cite{shibaI} 
and in post-Newtonian studies.\cite{shibaPN} 
Hereafter, we refer to the gauge condition presented here as 
an ``approximate minimum distortion'' (AMD) gauge. 

The above described 
AMD gauge condition is expected to have the following 
properties: 
\begin{itemize}
\item In a spherical symmetric spacetime with 
conformal flat initial conditions,
it coincides with Smarr-York's MD gauge condition 
(Eq. (\ref{SMMD})) as well as with the MD gauge condition 
presented in this paper (Eq. (\ref{MDnew})), 
because in this case, $\tilde \gamma_{ij}=\delta_{ij}$ and 
$\pa_t {\tilde\gamma_{ij}}=0$. 
%%%%%%%%%%%%%%%%%%%
\item Using a post-Newtonian approximation as 
a guide, we can see that the difference 
between our AMD gauge and the MD gauge condition 
defined by Eq. (\ref{MDnew})
appears, in general, at the third post-Newtonian 
order, because we neglect only 
coupling terms between $h_{ij}$ (or $f^{ij}$) 
and $\beta^k$.\cite{asada,ASF} 
Hence, this difference is in general small, and we may 
expect that 
the global change rate of $\tilde \gamma_{ij}$ is still 
suppressed and is always sufficiently small. 
%%%%%%%%%%%%%%%%%%%
\item As in the MD gauge conditions, 
in the AMD gauge condition, $h_{ij}$ approximately satisfies 
a transverse relation ($F_i=0$) everywhere except for a region 
near a rapidly rotating object 
as long as $F_i$ is sufficiently small at $t=0$ (see Appendix B). 
Thus, we expect that such gauge condition will prove useful for 
analyzing gravitational waves in the wave zone, in which 
$h_{ij}$ approximately satisfies a 
transverse-traceless (TT) condition, $h_{ij}\delta^{ij}=
h_{ij,k}\delta^{jk}=0$. 
\end{itemize}
In conclusion, the AMD gauge condition presented here 
possesses many of the merits of the MD gauge conditions, and 
it also results in equations which are much more easily 
solved numerically. 

\section{Analysis of gravitational waves} 

In this paper, we analyze gravitational waveforms 
as measured by an observer located near the outer boundaries 
of the computational domain. 
Along the $z$-axis, this can be done through the 
quantities 
\beqn
&&h_+ \equiv r(\tilde \gamma_{xx} - \tilde \gamma_{yy})/2,\\
&&h_{\times} \equiv  r \tilde \gamma_{xy}. 
\eeqn
Since we adopt our AMD gauge condition and 
consider initial conditions with $F_i=0$, $h_{ij}$ 
is approximately TT in the wave zone. 
As a result, $h_+$ and $h_{\times}$ are expected to be 
appropriate measures of gravitational waves emitted. (See 
\S 5 for a discussion of this.) 
The amplitude of gravitational waves will be 
largest along the $z$-axis as a result 
of our construction of initial data sets (see \S 5.1), 
and therefore $h_+$ and $h_{\times}$ 
should be regarded as 
the maximum amplitude of gravitational waves.

Gravitational waves are also measured through the gauge invariant 
Moncrief variables 
in flat spacetime, which are defined as follows:\cite{moncrief} 
First, we transform from the Cartesian 
coordinates to the spherical polar coordinates, 
and then we split $\gamma_{ij}$ 
as $\delta_{ij}+\sum_{lm} \zeta_{ij}^{lm}$, 
where $\zeta_{ij}^{lm}$ denotes 
\beqn
\zeta_{ij}^{lm}=&& \left(
\begin{array}{lll}
\displaystyle 
H_2 Y_{lm} & h_{1lm} Y_{lm,\theta}& h_{1lm} Y_{lm,\varphi}\\
*& r^2(K_{lm}Y_{lm}+G_{lm}W_{lm})&r^2G_{lm}X_{lm} \\
*& *&r^2\sin^2\theta(K_{lm}Y_{lm}-G_{lm}W_{lm}) \\
\end{array}
\right) \nonumber \\
&&+\left(\begin{array}{ccc}
0 &  -C_{lm} \pa_{\varphi} Y_{lm}/\sin\theta
& C_{lm} \pa_{\theta}Y_{lm}\sin\theta  \\
\ast & r^2D_{lm}X_{lm}/\sin\theta
             & -r^2 D_{lm}W_{lm}\sin\theta  \\
\ast & \ast & -r^2 D_{lm}X_{lm}\sin\theta \\
\end{array}
\right),
\eeqn
and $\ast$ the relation of symmetry. 
$H_{2lm}$, $h_{1lm}$, $K_{lm}$, $G_{lm}$, $C_{lm}$ and 
$D_{lm}$ are functions of $r$ and $t$, and are calculated 
by performing integrations over a two sphere of given radius 
(see a previous paper \cite{SN} for detail). 
$Y_{lm}$ is the spherical harmonic function, and 
$W_{lm}$ and $X_{lm}$ are defined as 
\beqn
&&W_{lm} \equiv \Bigl[ (\pa_{\theta})^2-\cot\theta \pa_{\theta}
-{1 \over \sin^2\theta} (\pa_{\varphi})^2 \Bigl] Y_{lm},\\
&&X_{lm} \equiv 2 \pa_{\varphi} \Bigl[ \pa_{\theta}-\cot\theta \Bigr] 
Y_{lm}. 
\eeqn
The Moncrief variables of even and odd parities are then 
defined as 
\beqn
&&R_{lm}^{\rm E}(t,r) \equiv 
\sqrt{2(l-2)! \over (l+2)!}
\Bigl\{ 4k_{2lm}+l(l+1)k_{1lm} \Bigr\}, \\
&&R_{lm}^{\rm O}(t,r) \equiv \sqrt{2(l+2)! \over (l-2)!}
\biggl(2{C_{lm} \over r}+r \pa_r D_{lm}\biggr),
\eeqn
where
\beqn
k_{1lm}&& \equiv K_{lm}+l(l+1)G_{lm}+2r \pa_r G_{lm}-2{h_{1lm} \over r},\\
k_{2lm}&& \equiv {H_{2lm} \over 2} - {1 \over 2}{\pa \over \pa r}
\Bigl[r\{ K_{lm}+l(l+1)G_{lm} \} \Bigr].
\eeqn
Note that using the above variables, 
the energy luminosity of gravitational waves can be calculated as 
\beq
{dE \over dt}={r^2 \over 32\pi}\Bigl[
(\pa_t R_{lm}^{\rm E})^2+(\pa_t R_{lm}^{\rm O})^2 \Bigr]. 
\label{dedt}
\eeq
In this paper, we consider only the even parity 
Moncrief variables of $l=|m|=2$, 
which are expected to have the largest amplitude.

\section{Numerical study}

We have carried out numerical simulations of the 
merger between two identical relativistic clusters 
of collisionless particles 
for several sets of initial conditions. In the following, 
we present the results obtained and demonstrate that the AMD gauge 
condition discussed in \S 3 is robust and allows us to 
perform simulations for a time much longer than the dynamical 
time scale and to extract gravitational waveforms. 

\subsection{Initial conditions}

For simplicity, we use conformally flat initial conditions. 
In this case, we only need to solve the Hamiltonian 
and momentum constraint equations, in addition to the equations for 
$\alpha$ and $\beta^k$. 
The constraint equations are written as \cite{gw3d}
\beqn
&& \Delta \psi = -2\pi E \psi^5 -{1 \over 8}
\hat A_{ij} \hat A^{ij}\psi^{-7} \label{ham2} \\
&& \hat A^{~j}_{i~,j} = 8\pi J_i \psi^6,\label{mom2}
\eeqn
where $\hat A_{ij} \equiv \psi^6 \tilde A_{ij}$. As discussed in a 
previous paper,\cite{gw3d} we write $\hat A_{ij}$ as 
\beq
\hat A_{ij}=W_{i,j}+W_{j,i}-{2 \over 3}\delta_{ij} W_{k,k},
\label{eq53}
\eeq
with 
\beq
W_i={7 \over 8}B_i - {1 \over 8}\Bigl(\chi_{,i}+B_{k,i} x^k\Bigr). 
\label{eq54}
\eeq
Then, the equations for $B_i$ and $\chi$ are derived from Eq. 
(\ref{mom2}) as 
\beqn
&&\Delta B_i = 8\pi J_i \psi^6 = 8\pi \rho_* u_i, \label{beq}  \\
&&\Delta \chi= -8\pi J_i x^i \psi^6 = -8\pi \rho_* u_i x^i.
\label{chieq}
\eeqn
As a result of Eqs. (\ref{eq53}) and (\ref{eq54}), 
Eqs. (\ref{ham2}), (\ref{beq}) and (\ref{chieq}) are 
regarded as constraint equations to be solved. 
The strategy for setting up initial conditions is as follows: 
(1) we choose $\rho_*(x,y,z)$ and $u_i(x,y,z)$; 
(2) we solve Eqs. (\ref{beq}) 
and (\ref{chieq}), from which $\hat A_{ij}$ is subsequently 
obtained; 
(3) we solve Eq. (\ref{ham2}). 

The rest mass density profile $\rho_*(x,y,z)$ of each cluster is 
simply chosen to be a parabolic function of the type 
\beq
\rho_*(r_{\pm})=\left\{
\begin{array}{ll}
\displaystyle \rho_0 
\Bigl( 1 -{r_{\pm}^2 \over r_0^2}\Bigr) & r_{\pm} < r_0, \\
0 & r_{\pm} \geq r_0, \\
\end{array}
\right.\label{rhosph}
\eeq
where $\rho_0$ and $r_0$ are constants, and 
$r_{\pm}$ denote coordinate distances from 
the center of each cluster which is located 
at $(\pm r_c,0,0)$ with $r_c>r_0$. 

As one model, we choose the velocity field $u_i$ for $r_{\pm} < r_0$ as 
\beq
u_i(x,y,z)=u_i^{\rm orb}=
\Bigl(-0.01{|x| \over x} V,~ 
{|x| \over x} V,~ 0\Bigr),\label{kepv}
\eeq
where 
\beq
V=\omega_0 \sqrt{M_*^{\rm one} \over 4r_c},
\eeq
and $M_*^{\rm one}=8\pi\rho_0 r_0^3/15$ 
is the total rest mass of one cluster. 
When we set the velocity field as Eq. (\ref{kepv}), we always 
refer to the model of the initial condition 
as Kepler model. In this model, 
we choose $\omega_0=1$ (near Kepler angular velocity case) 
and $0.5$. 
We also choose a nearly corotating velocity field as 
\beq
u_i(x^i)=u_i^{\rm orb}=
\Bigl(-0.01{|x| \over x} V-{y \over r_c} V,~ 
{x \over r_c} V,~ 0\Bigr).\label{corv}
\eeq
In this case, we set $\omega_0=1$. 
We refer to the model of the initial 
condition as corotating model. 

Once $\rho_*$ and $u_i$ are given, 
we can solve the constraint equations, as well as calculate 
the initial $z$-component of the 
total angular momentum of the system  
\beq
J=\int d^3x \rho_* (xu_y-yu_x)=j(\infty) . 
\eeq

Since we now consider the case in which each cluster is 
in near equilibrium, we must determine an appropriate 
velocity for each particle in addition to the orbital velocity. 
The method to determine the velocity 
$(u_i)_a$ for the $a$-th particle is as follows. 
First, we numerically solve equations for an equilibrium 
state of a single spherical cluster assuming that $\rho_*$ is 
given by the parabolic function as 
Eq. (\ref{rhosph}). The equilibrium state is easily 
obtained, as we have done in a previous paper.\cite{gw3d} 
Once we obtain an equilibrium state of a 
spherical cluster, the square of the velocity field, 
$u_i^{\rm one}u_i^{\rm one} \equiv \ell^2$, is obtained as 
a function of the coordinate distance from the center. 
Then, we assign the velocity of the $a$-th particle at 
$(x_a, y_a, z_a)=(x_p\pm r_c, y_p, z_p)$ as 
\beqn
&&(u_x)_a= \ell(r_p)\biggl({x_p z_p \over r_p \sqrt{x_p^2+y_p^2}}\cos\zeta
-{y_p \over \sqrt{x_p^2+y_p^2}} \sin\zeta \biggr)+u_x^{\rm orb},
\nonumber \\
&&(u_y)_a=\ell(r_p)\biggl({y_p z_p \over r_p \sqrt{x_p^2+y_p^2}}\cos\zeta
+{x_p \over \sqrt{x_p^2+y_p^2}} \sin\zeta \biggr)+u_y^{\rm orb},
\nonumber \\
&&(u_z)_a=-\ell(r_p){\sqrt{x_p^2+y_p^2} \over r_p} \cos\zeta,
\eeqn
where $r_p^2=x_p^2+y_p^2+z_p^2$, and random numbers 
of uniform deviation between $0$ and $2\pi$ are 
assigned for $\zeta$. 
Namely, the velocity of each particle is written as a 
linear combination of the velocity obtained in constructing 
a single equilibrium spherical cluster and of the orbital velocity. 
In order to guarantee that the velocity field is given by 
$u_i=u_i^{\rm orb}$ at any point, 
we have distributed particles in an appropriate manner. 
Hence, in the limit $r_c \rightarrow \infty$, 
each cluster is in a true equilibrium. 

In addition to the constraint equations, we also need to solve 
equations for $\alpha$ and $\beta^k$, where for the latter, 
we use the prescription discussed in the previous section. 
Since we require $\alpha$ to initially satisfy the maximal 
slice condition, $K=0=\pa_t K $, we solve the equation
\beq
\Delta (\alpha\psi) = 2\pi \alpha \psi^5 (E+2 S_k^{~k})
+{7 \over 8}\alpha \psi^{-7}\hat A_{ij} \hat A^{ij} ,
\eeq
after we have a solution for $\hat A_{ij}$ and $\psi$. 

\subsection{Numerical results}

All simulations were carried out fixing parameters as 
$r_0=1$ and $r_c=1.1r_0$, but using three different values 
for the total rest mass 
$M_*=2M_*^{\rm one}$, $r_0/6$, $r_0/4$ and $r_0/3$ 
(i.e., changing $\rho_0$). 
We note that in all cases, each spherical star 
is stable in isolation. 
Only when $M_*$ is greater than $\sim r_0/2.5$, 
a spherical star becomes unstable in isolation for the 
density profile given by Eq. (\ref{rhosph}), because 
some orbits near the surface of the cluster are unstable. 

We typically use a uniform grid with $(153, 77, 77)$ 
points and covering a coordinate domain with extents 
$-7.6r_0 \leq x \leq 7.6r_0$ and 
$0 \leq y, z \leq 7.6r_0$ (i.e., 
$\delta x=\delta y=\delta z =0.1r_0$). 
In this case, 
radius of each cluster is covered by $10$ grid points initially. 
We typically take the particle number $N$ to be $10^5$ 
\footnote{For some cases, we performed simulations changing 
$N$ and the grid spacing, and found that the numerical results 
changed somewhat in particular after the collision of two clusters. 
This is due to the different magnitude of the 
fluctuations caused by the discreteness of the particle locations. 
However, the global features discussed here did not change 
considerably.}. 
Hereafter, all quantities are given in units $r_0=1$ 
(and $G=1=c$). 

\subsubsection{Products of the merger}

In Figs. 1--3, we show snapshots of the particle positions at 
selected times for Kepler models with 
$M_*/r_0=1/6$, 1/4 and 1/3, respectively. 
(We list several characteristic quantities of the initial 
conditions in Table I.) In the first two cases, the final 
configuration consists of a new nearly axisymmetric, 
rotating cluster, while in the third case 
a black hole appears to have formed. In the last case, 
the simulation could not be continued until 
we determined the apparent horizon because the accuracy became 
insufficient. However, we presume that a 
black hole was formed, because the merged object was 
very compact and $\alpha$ and $\phi$ became, respectively, small 
($\sim 0.1$) and large ($\sim 0.7$) when the simulation 
was terminated. 
The reason that we had to terminate the simulation before we 
determined the apparent horizon is probably 
related to our AMD gauge, 
which seems inappropriate when applied to a 
problem in which black hole formation proceeds and the 
forming region is not well resolved. 
We discuss this point further below. 

For the cases in which $M_*/r_0=1/6$ and 1/4, 
we stopped our simulations at $t \sim 85r_0$ 
when the new cluster seems to settle down to a 
nearly equilibrium state. 
For $M_*/r_0=1/6$, 
the core within $r < 2r_0$ in the final state comprises 
about $82\%$ of the total rest mass (i.e., $m_*(2r_0)/M_* \simeq 0.82$) 
and about $48\%$ of 
the initial total angular momentum (i.e., $j(2r_0)/J \simeq 0.48$). 
For $M_*/r_0=1/4$, we have $m_*(2r_0)/M_* \simeq 0.83$ and 
$j(2r_0)/J \simeq 0.61$, respectively. 
It appears, therefore, that the final configuration 
consists of dense,  rapidly 
rotating clusters surrounded by a halo. 
The reason that $j(2r_0)/J$ for $M_*/r_0=1/6$ is 
smaller than that for $M_*/r_0=1/4$ is probably related to 
the fact that in the first case, 
self-gravity is weaker, and 
particles of large angular momentum can easily diffuse outward. 

As shown in Figs. 1--3, 
the merging process toward the final state 
considerably depends on the initial compaction parameters. 
For example, for $M_*/r_0=1/6$, 1/4, and 1/3, 
the merging starts after about one, 
three quarter and half an orbital periods, 
respectively. Also, for $M_*/r_0=1/6$, 
the merged object initially has a triaxial ellipsoidal shape 
which then gradually settles down to a spheroidal object forming 
spiral arms and transporting the angular momentum outward. 
On the other hand, 
for $M_*/r_0=1/4$ and 1/3, the merging proceeds more quickly and 
violently, forming 
a very high density object or a black hole around the mass center. 

Since we did not prepare a quasi-equilibrium configuration 
as initial conditions for the mergers, 
it is difficult to present a rigorous interpretation of the 
results. However, we believe that 
the different types of behavior at the initial stages 
of the merger can be related to one of the three following 
factors. $(a)$ In all cases, 
each cluster in a binary is tidally elongated along 
the axis which connects the two cluster centers just after the 
simulation starts. 
As a result of the elongation, each cluster 
in the binary acquires a quadrupole moment, which makes the attractive 
force between the two clusters stronger and can trigger the merging. 
Since the elongation may be larger for binaries of 
larger compaction due to a general relativistic effect, 
the attractive force becomes stronger 
for clusters of larger initial compaction. 
$(b)$ Gravitational radiation reaction, 
which accelerates the merging, should be stronger for 
more compact binaries. This can be 
clearly deduced from the ratio of the coalescence time 
to the orbital period, which can be 
approximately written as \cite{ST}
\beq
R_{\tau} \equiv {5 \over 128\pi}\biggl({2r_c \over M_*}\biggr)^{5/2}. 
\eeq
$R_{\tau}$ is $\sim 1.4$ for $M_*/r_0=1/3$, while it is 
$\sim 8$ for $M_*/r_0=1/6$. Thus, gravitational radiation 
can affect the evolution of the compact binary more efficiently. 
$(c)$ For $M_*/r_0=1/3$ and 1/4, the ratio of the initial 
orbital radius to the gravitational mass $M$ is $\sim 7$ and 9, 
respectively, so that the orbital radius may be 
as small as the radius of the inner most stable orbit 
of the system. 
This implies that, for $M_*/r_0=1/3$ and 1/4, 
a rapid transition from a stable near-circular orbit to 
a plunging orbit probably takes place 
when the orbital radius slightly decreases. 
On the other hand, for $M_*/r_0=1/6$, 
this is never the case. 

Note that all the above factors are 
general relativistic. Therefore, 
although we cannot make a firm statement on the origin of 
the different types of behavior for different initial compaction, 
it is almost certain that a general relativistic effect, 
which is never observed in Newtonian simulations, is crucial 
in these simulations. 

The evolution in the late phase of the merger 
is also different for the three models. 
In particular, for $M_*/r_0=1/4$, we find an interesting 
feature which 
appears to be a peculiarity of 
the merger between collisionless clusters. 
In this case, the 
two clusters merge to form a very high density 
cluster in the early phase ($t\sim 30r_0$). 
Then, a large fraction of particles 
expands outward ($t\sim 30r_0-50r_0$), and 
the central density decreases. Because the system is 
strongly bound, the particles eventually converge in the 
central region ($t> 50r_0$) to produce a final compact cluster. 
A possible explanation for the outward motion for $t \sim 30-50r_0$ 
can be found in the peculiar nature of a collisionless 
system, which does not possess any efficient mechanism 
of conversion of the kinetic energy into internal energy. 
In contrast, in the merger between fluid stars, 
a shock is formed, and the kinetic energy may be converted to 
the thermal energy. Consequently, most of the mass (or energy) 
in the system may be trapped near the mass center.

For $M_*/r_0=1/4$, 
we have also performed a simulation using 
the zero shift gauge condition 
($\beta^k=0$). In this case, the large coordinate distortion 
leads to increasing large values of $\tilde \gamma_{ij}$ and 
produces very inaccurate results in a time 
$\sim 1/3$ orbital period. 
We show $h_{xx}$ and $h_{yy}$ 
in the equatorial plane for 
$\beta^k=0$ and $M_*/r_0=1/4$ 
at selected times in Figs. 4(a) and (b), respectively. 
It is found that the maximum (minimum) value 
monotonically increases (decreases), and near $t \sim 13r_0$ it 
becomes $\sim 9$ ($-0.5$) for $h_{xx}$ and 
$\sim 2.4$ ($-0.5$) for $h_{yy}$. 
In contrast, in the simulations with our AMD gauge, the 
maximum (minimum) value of $h_{xx}$ 
and $h_{yy}$ does not monotonically 
increase (decrease) (see Figs. 4 (c) and (d)). 
Since we start the simulation with 
a conformal flat initial condition, 
it initially increases or decreases monotonically, but 
the maximum (minimum) stops increasing (decreasing) 
in about one orbital period when they have reached 
$\sim \pm 0.05$. 
This behavior is a clear indication that the AMD gauge condition 
provides the required suppression of 
the coordinate distortion. 

Note that the magnitudes of 
$h_{ij}$ are roughly consistent with 
$h_{ij}$ being a second post-Newtonian quantity, i.e., of 
$O[(M_*/r_0)^2]$. This confirms that 
our AMD gauge condition is actually a valid approximation 
of the MD gauge conditions, because the coupling terms 
between $h_{ij}$ and $\beta^k$ are sufficiently small. 
Since $h_{ij}$ does not 
monotonically increase nor decrease and 
the absolute value of each component remains small,
simulations can be stably and accurately 
performed for a time much longer than the dynamical time scale 
using our AMD gauge condition.

In Figs. 5 and 6, we show snapshots of the particle positions at 
selected times for corotating models of 
$M_*/r_0=1/4$ and $1/3$. These figures 
should be compared with Figs. 2 and 3, respectively. 
As in the Kepler case, the final product of 
the less compact binary is a 
rotating cluster, while that of the more compact binary 
appears to be a black hole. In the case $M_*/r_0=1/3$, 
again the simulation could not be continued until the apparent 
horizon was determined. 

By comparing Figs. 2 and 5, we find that 
the merging process in corotating models 
is different from that occurring in Kepler models, and 
this is particularly evident for 
the case $M_*/r_0=1/4$. In general, in the cases of 
corotating initial conditions, 
the outer part of the binary has more angular momentum 
than that for the Kepler model, so that the centrifugal force is 
stronger. Hence, the merging proceeds mainly in the inner region 
of the system, 
and particles in the outer region are not involved. 
Since the two clusters do not collide very quickly and 
the merging proceeds gradually, 
the resulting object does not expand outward significantly 
after the merger when compared with the Kepler case. 
The product of the merger settles down to a spheroidal object 
more gradually, 
forming spiral arms in the outer region. In this case, 
$m_*(2r_0)/M_* \simeq 0.79$ 
and $j(2r_0)/J \simeq 0.48$ 
in the final stage at which we terminated computation, i.e.,  
at $t \sim 80r_0$. The fact that 
the final value of $j(2r_0)/J$ is smaller than 
that for the Kepler model with the same initial compaction 
results from the initial corotating  
velocity field, in which the outer parts have 
a large angular momentum 
and are therefore able to diffuse outward more easily. 

In order to clarify the reason that both our AMD gauge and 
the MD gauge conditions are not 
well suited to study the formation of a black hole, 
we also performed simulations for a Kepler model of 
$\omega_0=0.5$ with $M_*/r_0=1/3$. 
In this case, the angular momentum is not 
large enough to counteract the gravitational force between the two 
clusters, and they quickly merge into a black hole. 
In this simulation, we set $N=50000$, and the grid spacing was 
$\delta x =r_0/15$. 

As in the Kepler and corotating cases with $\omega_0=1$ and 
$M_*/r_0=1/3$, 
the simulation using the AMD gauge condition 
terminated before an apparent horizon was formed. 
The reason for this seems to be the following: 
Using a MD gauge condition or our AMD gauge 
condition with $K=0$, 
Eq. (\ref{md2}) can be roughly approximated as 
\beq
\Delta \beta^i_{~,i} \sim 12\pi J_{i,i}~~~. 
\eeq
During a gravitational collapse we have 
\beq
J_r<0,~{\rm and}~J_{i,i} \sim {1 \over r^2}\pa_r (J_r r^2) < 0,
\eeq 
where $J_r$ denotes the radial component of $J_i$ and 
we assume that the matter collapses mainly in the radial direction. 
Thus, in the central 
region where the black hole formation takes place, 
the divergence of 
$\beta^i$ is positive, i.e.,  
\beq
\beta^i_{~,i} >0 ~{\rm and}~ \beta^r >0. 
\eeq
Clearly, 
the condition $\beta^r>0$ implies that coordinates spread 
radially outward; 
i.e., the physical size between two neighboring grids increases and 
the numerical resolution deteriorates.\cite{PST} 
At $r=0$, $\phi$ increases rapidly 
when the collapse begins (see the solid line 
in Fig. 7); for $t \sim 14r_0$, $\phi \sim 0.8$ while 
it was $\sim 0.13$ at $t=0$. 
As a result of this variation in $\phi$, 
the physical separation 
between two neighboring grid points around $r=0$ increases 
by a factor of $e^{1.4}\sim 4$. 
If we require the same accuracy as that at $t=0$, 
we have to reduce the grid spacing 
$\delta x$ by a factor of $4$, which is not feasible  
in a 3D numerical simulation. 
This problem might be resolved with the help of 
an adaptive mesh refinement (AMR) technique.\cite{AMR} 
However, $\phi$ increases more and more 
as the collapse proceeds, so that even an AMR treatment might be 
insufficient without changing the gauge condition once a 
gravitational collapse begins. 
In this paper, thus, 
we propose a modified spatial gauge condition derived from 
our AMD condition as 
\beq
\beta^k=\beta^k_{\rm AMD}-{x^k \over r} \beta^r_{\rm AMD} f(r,t),
\eeq
where $\beta^k_{\rm AMD}$ is the shift vector determined 
from our AMD gauge condition, $\beta^r=x^k\beta^k_{\rm AMD} /r$, and 
$f(r,t)$ is a function of $r$ which satisfies $f(r=0)=O(1)$ and 
$f(r=\infty)=0$. In this case, 
$\beta^r$ around a black hole forming region 
can be set to a small value. 
Although the distortion in $\tilde \gamma_{ij}$ due to 
the radial motion increases, the coordinate twisting 
due to the angular motion may still be suppressed. 
Furthermore, the TT property is preserved in the wave zone 
because $f(r,t)\rightarrow 0$ for $r\rightarrow \infty$. 
To validate this idea, 
we perform a simulation using the following $f(r,t)$ as 
an example, 
\beq
f(r,t)=f_0(t){1 \over 1+(r/3M_*)^6},
\eeq
where we choose $f_0$ to be
\beq
f_0(t)=\left\{
\begin{array}{ll}
\displaystyle 
1 & {\rm for}~\alpha(r=0)  \leq 0.5, \\
0 & {\rm for}~\alpha(r=0) > 0.5. \\
\end{array}
\right.\label{f0eq}
\eeq 

In Fig. 7, we show $\phi$ at $r=0$ for $f_0$ of Eq. (\ref{f0eq}) 
(the dashed line) as well as for $f_0=0$ (the solid line). 
The behavior of $\phi$ clearly shows that the coordinate spreading is 
suppressed after we turn on a non-zero $f_0$. 
In Fig. 8, we also show the root mean square 
of the coordinate position for particles defined as 
\beq
r_{\rm rms}\equiv \sqrt{{1 \over N} \sum_{i,a} x^i_a x^i_a}. 
\eeq
Although the physical results of these simulations should be 
identical, the time evolution of 
$r_{\rm rms}$ in each simulation should be 
different when different gauge conditions are adopted. 
Since the coordinate position of each particle 
becomes smaller when the radial coordinate spreads outward, 
we expect $r_{\rm rms}$ to be smaller for a smaller $f_0$. 
Figure 8 clearly reflects this property and 
that the non-zero $f_0$ suppresses the coordinate spreading effect. 

In Figs. 9 and 10, we also show snapshots of 
particle positions at selected times for $f_0=0$ and the 
non-zero $f_0$, respectively. 
We find that in the $f_0=0$ case and at late time, 
particles are concentrated 
in very narrow regions around the origin and, as a consequence, 
the resolution becomes insufficient 
for $t \sim 14r_0$. On the other hand, 
in the non-zero $f_0$ case, the coordinate spreading was not 
appreciable, at least up to the formation of an 
apparent horizon, and the simulation continued 
after the formation of the apparent horizon and for a long enough 
time that the horizon stretching became dominant and spoiled 
the numerical accuracy. 
In conclusion, we can mention that in addition to our 
AMD gauge condition, it may be helpful to 
incorporate an appropriate non-zero function $f(r,t)$ in the final 
phase of the merger just before the formation of a black hole. 

It is important to stress that the present strategy to 
compensate for the limitation of MD-type gauge conditions is 
well suited to the problem under investigation and is 
not the only possible solution. In a different scenario, 
different solutions might turn out to be more effective. 
Consider, for example, the case in which a rapidly rotating 
object collapses and a disk is formed prior to the 
formation of a black hole. 
In this case, it would probably be more appropriate to 
suppress $\beta^z$ first, 
and then, $\beta^R[\equiv (x\beta^x+y\beta^y)/\sqrt{x^2+y^2}]$ 
only when the black hole formation starts. It is evident 
that more detailed studies 
on the choice of the gauge condition 
suitable to study the formation of a black hole are necessary. 

\subsubsection{Gravitational waveforms}

In Figs. 11 and 12, we show $h_+/(M_*^2/r_0)$ and 
$h_{\times}/(M_*^2/r_0)$ at 
$z_{\rm obs}=7.5r_0$ (solid lines) and $6r_0$ (dashed lines) 
for the Kepler and corotating models with $M_*=r_0/4$, respectively. 
We plot these lines as a function of $(t-z_{\rm obs})/r_0$. 
Hence, if $h_+$ and $h_{\times}$ behave as 
solutions of the wave equations near the outer boundaries, 
the two lines should approximately agree. 
Note that gravitational waves begin to reach the observer at 
$t-z_{\rm obs} \simeq 0$, 
so that the waveforms shown for $t-z_{\rm obs}<0$ are meaningless 
as the coalescence waveforms. 

It should be mentioned that the expected wavelength of 
gravitational waves in the initial phase of the merger 
is approximately $T/2 \simeq 20r_0$, where 
$T$ is an approximate orbital period 
(see Table I for definition of $T$). In principle, 
the asymptotic waveforms should be extracted 
in the wave zone, and therefore for $z_{\rm obs} > 20r_0$. 
Nevertheless, the waveforms shown here appear to constitute 
a fair description of the asymptotic waveforms due to 
the following reasons: 
$(i)$ The solid and dashed lines agree well in both 
cases throughout the computations, 
so that $h_+$ and $h_{\times}$ actually propagate at the 
speed of light. 
$(ii)$ The wavelength in the initial phase of the merger 
approximately agrees with that expected from the 
initial orbital period. 
$(iii)$ The amplitude of $h_+$ and $h_{\times}$ in the 
initial phase roughly 
agrees with an analytical estimate using the quadrupole 
formula for two point masses in a circular orbit, i.e., 
$M_* (4\pi r_c/T)^2=M_*^2/2r_c$.\cite{ST} 
$(iv)$ The waveforms agree fairly well with those 
expected from Figs. 2 and 5; i.e, in their initial merging phases, 
the amplitudes of gravitational waves are the largest 
because the systems are highly non-axisymmetric, 
but once they settle down to form an elliptical 
merged object, the amplitude gradually decreases.

The gravitational waveforms in Figs. 11 and 12 also reflect 
the difference in the merger process for the Kepler and 
the corotating models. 
In the Kepler case, the merging quickly proceeds to form 
a new, high density rotating object 
resulting in a rapid and large increase in the amplitude of 
gravitational waves. 
In the corotating model, on the other hand, 
the merging proceeds less rapidly, and the resulting 
amplitude of gravitational waves is smaller. 
Also, in the Kepler case, a high density and rapidly 
rotating ellipsoidal object 
is formed after the merger, and it is responsible for the 
emission of high frequency gravitational waves. 
The results presented here suggest that waveforms of gravitational 
waves from the merger of stellar clusters may be sensitive to the 
internal velocity fields that two clusters possess 
before the merger, and could be used to extract 
astrophysical information. 

In both the Kepler and the corotating cases, 
the maximum values of $h_+$ and $h_{\times}$ are roughly 
$\sim 0.08M_*(4M_*/r_0)$ and the wavelength is 
$\sim 10r_0(r_0/4M_*)^{1/2}$, as expected from the 
quadrupole formula. If similar mergers 
between highly relativistic clusters actually occurred in the early 
universe, the amplitude of gravitational waves could be 
\beq
\sim 10^{-18}\biggl({4000{\rm Mpc} \over r}\biggr)
\biggl({M_* \over 10^6M_{\odot}}\biggr)
\biggl({4M_* \over r_0}\biggr),
\eeq
with the frequency $\sim 10^{-2}(10^6M_{\odot}/M_*)
(4M_*/r_0)^{3/2}$Hz, 
where $M_{\odot}$ denotes the solar mass. 
If mergers between highly relativistic star clusters 
of mass $\sim 10^6M_{\odot}$ and radius $\sim 10M_*$ 
occurred in the early universe, it should be possible to 
detect emitted gravitational waves 
by the planned gravitational wave detectors in space.\cite{LISA} 

In Figs. 13 and 14, 
we show $r R_{2,2\pm}/r_0$ as a function of $(t-r_{\rm obs})/r_0$ 
for the Kepler and the corotating models 
of $M_*=r_0/4$, where we define 
\beqn
R_{2,2+}\equiv {R_{2,2} + R_{2,-2} \over \sqrt{2}},\\
R_{2,2-}\equiv {R_{2,2} - R_{2,-2} \over \sqrt{2}i}.
\eeqn
The solid and dashed lines denote the waveforms extracted at 
$r_{\rm obs}=6.5r_0$ and $7.3r_0$, respectively.

We find that for $r R_{2,2\pm}$, 
the solid and dashed lines do not coincide well 
compared with those for $h_+$ and $h_{\times}$. 
In particular, for $t-r_{\rm obs} < 10r_0$, 
in which case the variation time scale of the system is 
relatively long, the coincidence 
is not good. The disagreement is probably due to the fact that 
the main contribution to $R_{2,2\pm}$ comes 
not only from the $O(r^{-1})$ part of $\tilde \gamma_{ij}$ but also 
from the $O(r^{-3})$ part of $\exp(\phi)$, which 
can be approximated for large $r$ as
\beq
\exp(\phi)=1+{M \over 2r}+{3\bI_{ij} x^i x^j \over 4r^5} + 
O(r^{-5}),
\eeq
where $\bI_{ij}$ denotes the trace free part of a 
quadrupole moment. An order of magnitude estimate for 
$l=|m|=2$ modes of 
$h_{ij}$ and $\exp(\phi)$ is given by 
\beq
h_{ij} \sim {1 \over r}{d^2\bI_{ij} \over dt^2},~{\rm and}~
\exp(\phi) \sim {\bI_{ij} \over r^3}.
\eeq
As a result, the ratio of the magnitude of 
$h_{ij}$ to that of $\exp(\phi)$ is $\sim (\omega r)^2$, 
where $\omega$ denotes a characteristic angular frequency of 
gravitational waves. In order to extract gravitational 
waves (i.e., the contribution from $h_{ij}$) clearly, 
$\omega r$ should be much larger than unity. However, 
for $t-r_{\rm obs}\sim 0$, 
we extract $R_{2,2\pm}$ for $ r \sim \omega^{-1}$, and 
the magnitudes of the two contributions are roughly equal. 
For this reason, the solid and dashed lines 
do not coincide well, 
%%\footnote{For simulations with smaller compaction, 
%%the contribution from $\exp(\phi)$ is larger because 
%%$\omega$ is smaller. Hence, the situation becomes even worse.}. 
and $R_{2,2\pm}$ for $t-r_{\rm obs} < 10r_0$ 
should not be regarded as accurate quantities for 
measuring gravitational waves. 

However,  when $\omega$ is sufficiently large, 
the coincidence between the solid and dashed lines 
is improved, and for 
$t-r_{\rm obs} \sim 10-40r_0$, the coincidence is fairly good. 
Assuming the quantities $R_{2,2\pm}$ describe 
gravitational waveforms correctly, 
we can calculate the total energy luminosity $\Delta E$ 
due to $l=|m|=2$ components by integrating $dE/dt$ from 
$t-r_{\rm obs}=10r_0$ to the end of the simulations using 
Eq. (\ref{dedt}). 
We obtain for the Kepler and corotating models, respectively, 
\beq
\Delta E \simeq 4.5 \times 10^{-3}~~{\rm and}~~
1.5 \times 10^{-3} M_*.\label{delE}
\eeq
For a binary composed of equal masses in a circular 
orbit of orbital radius $2r_c$, the energy radiated 
by gravitational waves in one orbital period is 
roughly given by the quadrupole formula as \cite{ST} 
\beq
\Delta E\sim {4\pi \over 5}\biggl( {M_* \over 2r_c}\biggr)^{7/2}M_*
=1.24\times 10^{-3} M_* \biggl({4M_* \over r_0}\biggr)^{7/2}
\biggl({1.1r_0 \over r_c}\biggr)^{7/2},
\eeq
which confirms the validity of the results obtained in 
Eq. (\ref{delE}). 
The reason that $\Delta E$ for the Kepler model is larger than that 
for the corotating model is that the merging 
proceeds more quickly and that 
a rapidly rotating ellipsoidal core of a higher density 
is formed in the initial stage of the merger. 
Irrespective of the used model, however, 
these results apparently indicate that a merger between 
two relativistic clusters is an efficient 
source of gravitational waves.

\section{Summary}

We have performed fully general relativistic simulations of 
the merger between relativistic clusters of collisionless particles 
adopting a new spatial gauge, 
the AMD (approximate minimum distortion) gauge 
condition. We have reached the following conclusions:
\begin{itemize}
\item By using the AMD gauge condition, it is possible to 
perform numerical simulations of coalescing 
binary clusters for a time much longer than 
the dynamical time scale of the system 
when the merger does not produce a black hole. Also, it is 
found that with this gauge condition, 
gravitational waveforms can be calculated fairly accurately. 
These results suggest that with the AMD gauge 
condition, the spatial coordinate distortion is suppressed 
to a level adequate to carry out stable and accurate 
simulations. 
%%%%%%%%%%%
\item Our AMD gauge condition, as well as a 
class of spatial gauge conditions similar to the MD 
gauge conditions, 
still has an undesirable, coordinate spreading property 
(i.e., $\beta^r>0$) when applying it to the 
study of black hole formation. 
As a result of this property, the spatial resolution around 
a black hole forming region rapidly tends to become insufficient.
%%%%%%%%%%%
\item However, if our AMD gauge condition is slightly modified 
during the process of black hole formation, 
it is still possible to carry out simulations 
up to the formation of an apparent horizon. 
Using this approach, we have shown that 
it is possible to perform 
numerical simulations over a sufficient long time scale from 
the early merging phase up to the formation of a black hole. 
%%%%%%%%%%%
\item In a merger between two highly relativistic 
star clusters with $M_*/r_0=1/4$ into a new cluster, 
the amplitude of gravitational waves has been estimated as 
$\sim 10^{-18}(M/10^6M_{\odot})$ 
at a distance of $\sim 4000$Mpc with a frequency 
$\sim 10^{-2}(10^6M_{\odot}/M)$Hz. 
Also, $\sim 0.5\%$ of the rest mass energy may be dissipated by 
gravitational waves in the final phase of the coalescence. 
\end{itemize}
\vskip 2mm

The spatial gauge conditions presented and investigated 
in this paper are useful for a wide variety of problems, 
and we expect that they 
will be widely exploited for performing simulations of 
coalescing binary neutron stars, which represent 
the most promising sources of gravitational waves 
for kilometer-size laser interferometers. 
When a general relativistic hydrodynamic code is installed 
to replace numerical code for collisionless particles, 
we will be able to perform the numerical simulations for 
such binary systems and follow them up to the formation of 
black holes or new rotating neutron stars. 
This work is now in progress, and 
in future papers we will present the numerical results.\cite{gr3d}

\vskip 3mm
\begin{center}
{\large\bf Acknowledgments}
\end{center}
\vskip 3mm

The author thanks T. Nakamura, M. Sasaki, S. Shapiro and 
L. Rezzolla for frequent 
helpful conversations. He also thanks L. Rezzolla for 
carefully reading the manuscript 
and for providing helpful comments. 
Numerical computations were performed on the FACOM VX/4R machine in 
the data processing center of the National Astronomical Observatory 
of Japan (NAOJ). 
This work was supported by a Grant-in-Aid (Nos. 08NP0801 and 
09740336) of 
the Japanese Ministry of Education, Science, Sports and Culture, 
and JSPS Fellowships for Research Abroad. 

\appendix

\section{Numerical treatment for the transport terms} 

For treating the transport terms in the evolution equations 
of geometric variables, we use a method similar to that 
adopted by Stark and Piran.\cite{SP} In the following, we demonstrate 
the treatment only in the $x$ direction, but the 
same operations are also carried out in the $y$ and $z$ directions. 
We also assume a uniform grid along the $x$ direction and 
denote the grid spacing and time step as $\delta x $ and 
$\delta t$. 

Evolution equations for the geometric variables 
may be written in the form 
\beq
(\pa_t - \beta^x \pa_x) Q=S,
\eeq  
where $Q$ denotes one of the geometric variables and $S$ 
denotes the source term. 
The solution for the finite difference equation 
is written 
\beq
Q^{n+1}_i=Q^{n}_i - {1 \over 2} \Bigl[ \nu (Q^n_{i+1}-Q^n_{i-1})
- |\nu|(Q^n_{i+1}-2Q^n_i+Q^n_i)\Bigr]+S^{n+1/2}_i \delta t,
\eeq
in the first order upwind scheme, and 
\beq
Q^{n+1}_i=Q^{n}_i-{1 \over 2}\Bigl[ \nu (Q^n_{i+1}-Q^n_{i-1})
-\nu^2(Q^n_{i+1}-2Q^n_i+Q^n_{i-1})\Bigr] + S^{n+1/2}_i \delta t,
\eeq
in the second order scheme. 
Here, $\nu=-(\beta^x)^{n+1/2}_i \delta t / \delta x$, 
and $Q^{n}_i$ denotes 
$Q$ at the $n$-th time step and $i$-th grid point. 
In this paper, we compute $Q^{n+1}_i$ as 
\beqn
Q^{n+1}_i=Q^{n}_i-{1 \over 2} \Bigl[ \nu (Q^n_{i+1}-Q^n_{i-1})
&&-(1-s_i)|\nu|(Q^n_{i+1}-2Q^n_i+Q^n_{i-1}) \nonumber \\
&&-s_i\nu^2(Q^n_{i+1}-2Q^n_i+Q^n_{i-1})\Bigr] + S^{n+1/2}_i \delta t.
~~~~~
\eeqn
Here, $s_i$ is a limiter function at the $i$-th grid point, chosen as
\beq
s_i=\left\{
\begin{array}{ll}
\displaystyle 
{2 \delta Q_i \delta Q_{i-1} +\epsilon \over 
\delta Q_i^2+\delta Q_{i-1}^2+\epsilon} & {\rm for}~
\delta Q_i \delta Q_{i-1}  \geq 0,\\
0 & {\rm for}~\delta Q_i \delta Q_{i-1} <0 , \\
\end{array}
\right.
\eeq
where $\delta Q_i=Q^n_{i+1}-Q^n_i$ and 
$\epsilon$ is an appropriately small constant.

\section{Behavior of $F_i$ in our AMD gauge}

Substituting Eq. (\ref{amdeq}) into Eq. (\ref{fijeq}), we obtain 
\beqn
(\pa_t - \beta^k \pa_k)F_i&& = 2\alpha \Bigl(f^{kj} \tilde A_{ik,j}
+f^{kj}_{~~,j} \tilde A_{ik}
-{1 \over 2} \tilde A^{jl} h_{jl,i}\Bigr)
+2f^{jk} \alpha_{,k} \tilde A_{ij} \nonumber \\
&&~~~ + \delta^{jl} \beta^k_{~,l}(h_{ij,k}+h_{ik,j})
+h_{il} \beta^l_{~,jk}\delta^{jk}+F_l \beta^l_{~,i}
+h_{jl}\beta^l_{~,ik}\delta^{jk} \nonumber \\
&& \hskip 1cm -{2 \over 3}(F_i \beta^l_{~,l}
+h_{ij} \beta^l_{~,lk} \delta^{jk}) \nonumber \\
&& \equiv S_i^{F}.\label{fijeq2}
\eeqn
Here $S_i^F$ is a nonlinear function and can be set to zero 
in the linear approximation. Hence, in this approximation, 
$F_i$ is always zero if it is zero initially. 
Even when we consider a linear perturbation 
in the Schwarzschild spacetime or the spacetime of a 
spherical star, $S_i^F$ is non-linear and $F_i$ is always 
zero if initially zero. However, in a spacetime of 
a rotating black hole or star, $h_{ij}$ ($f^{ij}$), 
$\beta^k$ and $\tilde A_{ij}$ 
appear at zeroth order, so $F_i$ also appears from zeroth 
order. Nevertheless, 
in the case that the rotation of the object is not very rapid, 
$h_{ij}$ ($f^{ij}$), 
$\beta^k$ and $\tilde A_{ij}$ are small, and hence $F_i$ can be 
considered to be small. 
Even in the case that the object is rapidly rotating, 
$h_{ij}$ ($f^{ij}$), $\beta^k$ and $\tilde A_{ij}$ quickly 
decay as $O(r^{-2})$, $O(r^{-2})$ and  $O(r^{-3})$, respectively, 
so that $S_i^F$ and $F_i$ 
can still be regarded as small quantities everywhere 
except near the rapidly rotating object. 
Furthermore, in the post-Newtonian approximation, 
$S_i^F$ contains at most third post-Newtonian quantities, 
because $h_{ij}$ ($f^{ij}$), $\beta^i$, and $\tilde A_{ij}$ 
can be regarded as the second, first, and 
first post-Newtonian quantities, 
respectively.\cite{asada,ASF} 
Thus, except in the highly relativistic region 
in which $h_{ij}$, $\beta^i$ and $\tilde A_{ij}$ become $O(1)$ 
or a relativistic region 
near the rapidly rotating relativistic objects, 
$S_i^F$ is guaranteed to be small, and 
hence, $F_i$ is expected to be small 
as long as it is 
small enough on an initial time slice (i.e., as long as 
we do not consider strange initial conditions in which 
$F_i$ is large initially).

%\clearpage
\vskip 5mm
\noindent 
{\bf Table I.~} The list of initial conditions and 
final states for simulations 
performed in \S 5. $M_*=2M_*^{\rm one}$, $M$, $J$ and $T$ denote 
the total rest mass, gravitational mass, total angular 
momentum and approximate initial orbital period 
($2\pi\sqrt{8r_c^3/M_*}$), respectively. 
All the quantities are given in units 
$r_0=1$ (and $G=1=c$). In the cases marked with $\dagger$, 
we could not determine the apparent horizon formation, 
but black holes seem to be formed. 

\vskip 5mm
\noindent
\begin{center}
\begin{tabular}{|c|c|c|c|c|c|c|} \hline
\hspace{2mm} $M_*$ \hspace{2mm} & \hspace{1mm} $M/M_*$\hspace{1mm} 
& \hspace{1mm} $J/M^2$ \hspace{1mm} & \hspace{2mm} $T$ \hspace{2mm}
& velocity field & Final state & Figures \\ \hline
1/6  &  $0.966$ & 0.974 & $50.2$ &Kepler($\omega_0=1$)~& 
rotating cluster&  1  \\ \hline
1/4  &  $0.952$ & 0.818 & $41.0$ &Kepler($\omega_0=1$)~& 
rotating cluster&  2 ,4, 11, 13  \\ \hline 
1/3  &  $0.941$ & 0.726 & $35.5$ &Kepler($\omega_0=1$)~&
black hole$^{\dagger}$ & 3 \\ \hline
1/4  & $0.955$ & 1.01 & $41.0$ & Corotation & 
rotating cluster & 5, 12, 14 \\ \hline
1/3  & $0.944$ & 0.894 & $35.5$ & Corotation & 
black hole$^{\dagger}$ & 6 \\ \hline
1/3 & $0.931$ &0.371 & $-$ & Kepler($\omega_0=0.5$)~&
black hole & 7--10 \\ \hline
\end{tabular}
\end{center}


\begin{thebibliography}{99}
\bibitem{asada} $h_{ij}$ and $\beta^i$ are the second and 
first post-Newtonian quantities, respectively, if $F_i$ is 
higher than third post-Newtonian order. In the present AMD gauge 
condition, $F_i$ can be regarded as a 
third post-Newtonian quantity (see Appendix B), 
so that $h_{ij}$ and $\beta^i$ are actually the second and 
first post-Newtonian order. 
This fact can be found by performing a post-Newtonian analysis 
in the (3+1) formalism as we have done in the next reference.\cite{ASF}
\end{thebibliography}
\end{document}